# Monetarization of the Feasible Operation Region based on a Cost-Optimal Flexibility Disaggregation

Marcel Sarstedt, Lutz Hofmann
Institute of Electric Power Systems, Electric Power Engineering Section
Leibniz Universität Hannover
Hanover, Germany
[surname]@ifes.uni-hannover.de

***Abstract*—Hierarchical multi-(voltage-)level grid control strategies are an appropriate design concept for the coordination of future TSO/DSO- and DSO/DSO-interactions. Hierarchical approaches are based on the aggregation of decentralized ancillary service potentials, represented by converter-coupled, communicable active and reactive power flexibility providing units (FPU) (e.g. wind turbines) at vertical system interfaces. The resulting PQ-polygon made available by the DSO for a potential request of ancillary service flexibilities by the TSO is called feasible operation region (FOR). A monetarization of the FOR is necessary for the implementation as operational degree of freedom within higher-level grid control. This paper presents an approach for the monetarization of the FOR in the context of a hierarchical multi-level flexibility market by a cost structure using metadata from population based optimization methods. Multiple FPU flexibility polygons at a single bus are aggregated for a reduction of the search space dimensions. The main contribution of the proposed method is the cost-optimal disaggregation of a flexibility demand to the single FPUs within the aggregated FPU by a mixed integer linear program (MILP). Therefore, a local flexibility market considering bids for the active and reactive power flexibilities by the FPUs stakeholders is assumed. The approach is applied within a case-study of the Cigré medium voltage system.**

***Index Terms*—Feasible Operation Region, Active Distribution Network, TSO/DSO-Cooperation, DSO/DSO-Cooperation**

## I. Introduction

The transition of the electric energy system leads to a massive integration of decentral energy resources (DER), especially to the distribution system level [1]. At the same time the contribution of conventional thermal power plants to the energy mix is decreasing in Germany, due to the coal and nuclear power phase-out and the priority feed-in of volatile renewables [1]. The ancillary service potentials of the transmission system operators (TSOs), guaranteeing a safe and reliable energy supply, are reduced. Additionally, there is a need for more control measures (e.g. active and reactive power redispatch) to avoid grid congestions and voltage band violations resulting from a local power imbalance [1].

A variety of system elements at the distribution system level are converter coupled and have an information and communication interface [2]–[7]. Thereby, they are flexibly controllable within the system operation of the corresponding distribution system operator (DSO). The possible adaptation of the active and reactive power supply of these flexibility providing units (FPU) can be described as polygon at the PQ-plane (see Fig. 1) [2], [3], [8]. The distribution system level transforms and becomes increasingly active. The flexibilities within active distribution networks (ADN) can be used for a change of the vertical interconnection power flows (IPF) at TSO/DSO- as well as DSO/DSO-interfaces according to the demand of the higher-level system operator and by this as additional ancillary service potential [9]–[14]. To coordinate the technical and organizational vertical interactions between the system levels an amendment of the regulatory framework at vertical system interfaces and the bilateral agreements between the TSOs and DSOs within novel multi-(voltage-)level grid control strategies is necessary [5], [7], [12], [15]. In literature especially hierarchical (aka. vertical distributed) multi-level grid control strategies are discussed for future TSO/DSO- and DSO/DSO-interactions [3], [12], [16]–[18]. In general, these approaches are based on a day-ahead or intraday, proactive determination of a feasible operation region (FOR) by the DSO by an aggregation of the flexibility potentials of the flexibility providing units connected to the distribution system level (see I. in Fig. 1) [2], [9], [11], [14], [19].

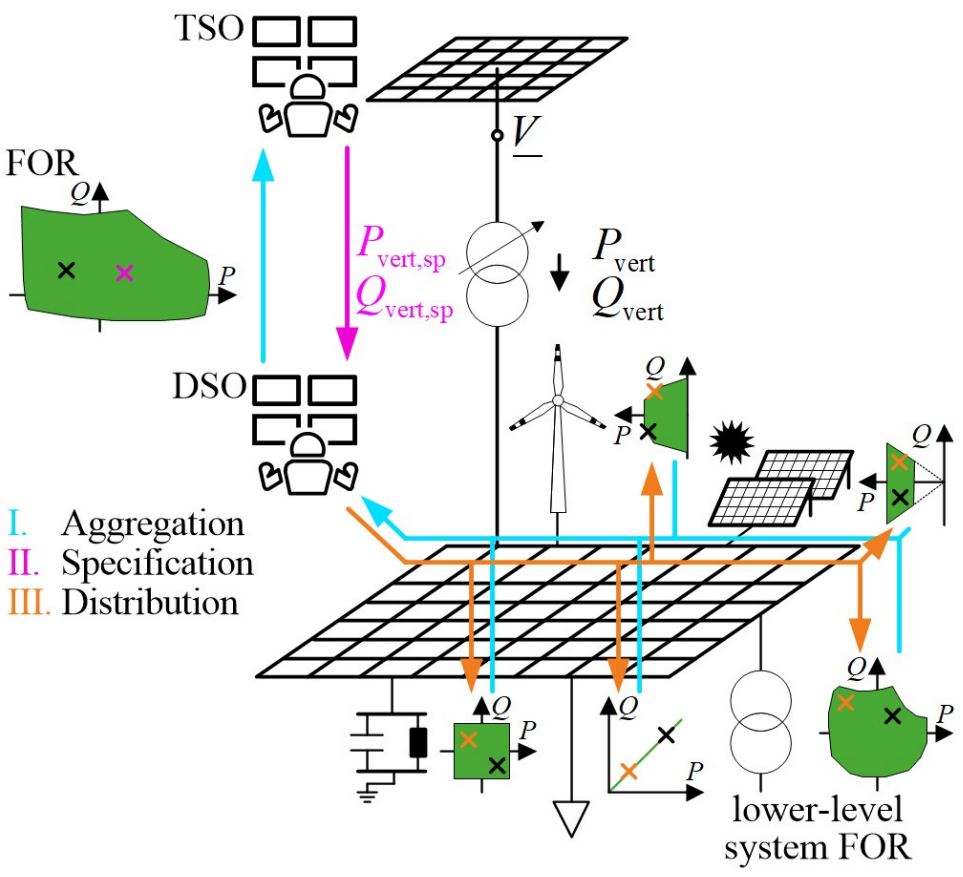

Figure 1. General process of hierarchical multi-level grid control strategies

The FOR describes the possible adaptation of the active and reactive IPFs ( $P_{\text{vert}}, Q_{\text{vert}}$ ) by the DSO within a PQ-plane. The next step is the specification of the flexibility demand ( $P_{\text{vert,sp}}, Q_{\text{vert,sp}}$ ) of the TSO from the distribution system level within the operational management of the TSO (see II. in Fig. 1). Finally, the higher-level flexibility demand is distributed to the FPUs within the operational management of the DSO (see III. in Fig. 1). Within the multi-level context, a cascading process based on bottom-up aggregation and top-down specification and distribution results [2], [3], [16].

## II. State of the Art of FOR Monetarization

The determination of the FOR at a single vertical system interface is based on an aggregation of the individual flexibility areas of the FPUs connected to lower-level grid. Rudimental approaches estimate the FOR (see Fig. 2a) or use the Minkowski sum (see [20], [21]) for the aggregation of polygonal flexibility areas of the FPUs, neglecting security constraints (e.g. voltage limits, maximum thermal currents) [22].

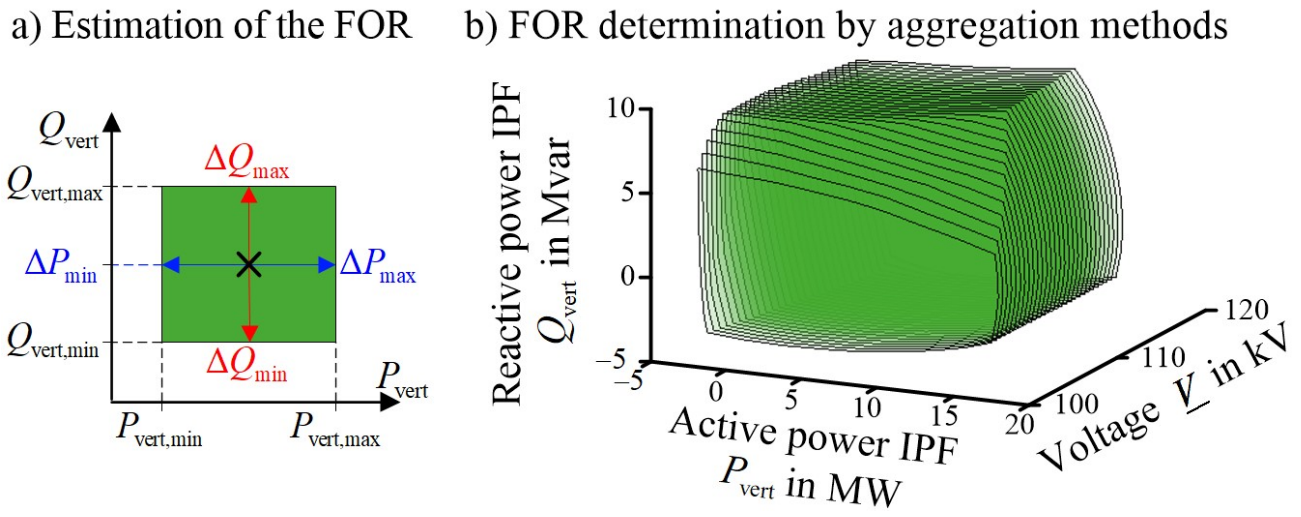

Figure 2. Exemplary FOR determined by a) simple estimation and b) advanced aggregation methods

In literature, two main categories of advanced aggregation methods also considering security constraints, the active and reactive power demand of the lower-level system as well as variations of the voltage at the interconnection bus $i$ as three-dimensional PQV-FOR (see Fig. 2b) are currently investigated. These are stochastic approaches using random search (RS) and optimization based approaches solving an adapted optimal power flow (OPF) problem (cf. [2], [23]).

RS aggregation methods (cf. [2], [14], [19], [24]–[26]) compute a variety of power flow calculations (e.g. Newton Raphson) for different constellations of operating points of the FPUs as Monte-Carlo scenarios. Each solution of a Monte-Carlo scenario represents a point at the PQ-plane which results in a point cloud. Disadvantages of RS methods are the long computation time and a challenging determination of the specific FOR edge especially in cases of non-convexities [14]. Straight forward RS approaches neglecting the covariance between the FPUs lead to a convolution of the results based on the central limit theorem of the probability theory [24], [27]. Novel RS approaches are using probability density functions for an appropriate sampling of the FOR [2], [28].

Advantages of RS approaches are the generation of metadata (e.g. flexibility provision per bus, bus voltages) for each point within the FOR (see Fig. 3a) and the simple consideration of non-convex FPU flexibility polygons [2], [23]. Another advantage is the good performance of the RS approaches for larger systems. The individual computation time for one power flow calculation increases proportional to the number of buses [2], [19]. Thereby, the computation time scales predominantly by the number of power flow calculations.

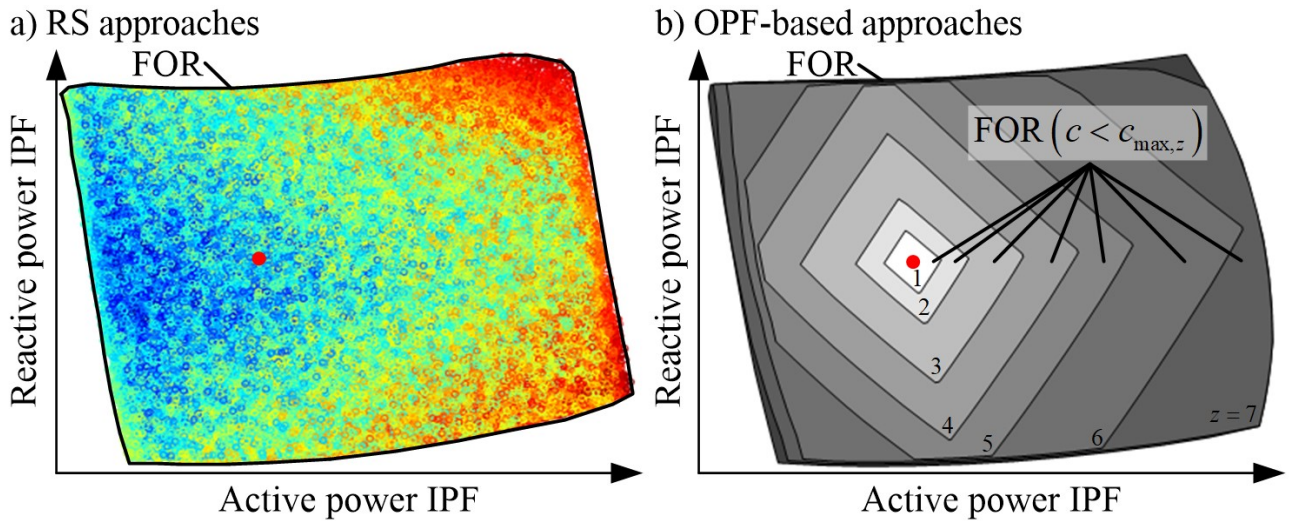

Figure 3. Exemplary monetarization of the FOR based on a) RS (blue = low service prize, red = high service prize) and b) OPF-based aggregation methods

OPF-based aggregation methods (cf. [9]–[11], [28]–[32]) sample the FOR edge in incremental steps. Therefore, at each sampling step an OPF with a specific constellation of the active and reactive IPFs is solved. The significance of the advantages of the OPF-based approaches depends on the specific optimization method for the solution of the OPF problem. Linear or quadratically constrained linear programming approaches (LP) determine the OPF edge in short computation times [8], [9], [23], [30], [33], [34]. The non-linear system behavior is only approximated which can lead to local convergence and by this to an over- or underestimation of the FOR compared to the FOR resulting from other optimization methods [2], [8]. This over- and underestimation of the FOR needs to be minimized to provide the higher-level system operator as much guaranteed flexibilities as possible in case of critical system states. The performance of LP scales proportionally for larger systems but the quality of the results decrease due to the increasing number of non-linearities [30], [35], [36]. Non-linear programming (NLP) approaches lead to an appropriate sampling of the FOR in short computation times also identifying non-convexities of the FOR [9]–[11], [23], [25]. The performance and the quality of the results of the NLP depends on the system size and the specific solver (e.g. interior point optimizer). With an increasing number of operational degrees of freedom and constraints also the possibility of local convergence increases [11].

The information of the cost structure of the FOR can be used by the higher-level system operator for an economical specification of the required flexibilities at the vertical system interconnection to the lower-level system within a hierarchical multi-level flexibility market [9], [14], [37], [38]. RS as well as OPF-based aggregation methods are compared with each other in literature (see [9], [14]) for the determination of the cost structure of the FOR. The description of the cost structure depends on the individual aggregation method. For RS approaches each point of the resulting point cloud at the PQ-plane can be simply monetarized based on the metadata received by the multiple power flow calculations (cf. [9]). The interpretation of the cost structure and the identification of zones with a uniform price is challenging (see Fig. 3a). The reason for this are multiple overlapping points resulting from various possible constellations of FPU flexibility provisions for a specific IPF. The resulting cost structure represents an estimation for the costs that a higher-level system operator can expect for a specific active and reactive power demand from the lower-level system.

Solely the prices of the FOR-edge are available after the sampling process for OPF-based approaches in contrast to RS approaches (cf. [14]). Multiple sampling processes that consider the compliance of specified prices as additional constraints are necessary to determine the cost structure within the FOR [14]. The cost structure is described by price zones $z$ whose individual contour lines represent a specific price $c_{\max,z}$ (see Fig. 3b) [14]. An advantage regarding RS approaches is the guarantee of minimum costs for the contour lines. Disadvantages are that the cost structure between two contour lines is not available. The computation time of OPF-based approaches is increased for an appropriate sampling of the cost structure.

Approaches using a metaheuristic (e.g. particle swarm optimization, genetic algorithm) combine the advantages of RS in generating metadata and the appropriate sampling of the FOR-edge of OPF-based methods [9], [23], [28]. In general, solving an OPF problem by metaheuristics is based on an evaluation of the objective function by a variety of power flow calculation, analogously to RS approaches. An iterative, algorithm-specific adaptation of the population is used for convergence in a solution of the OPF problem.

As typical for stochastic approaches, metaheuristics cannot guarantee global convergence or determine the remaining gap to the global optimum [39]. Nevertheless, the investigations in [23] show a good performance of the particle swarm optimization (PSO) in determining the edge of the FOR and a high quality of the results (e.g. size of the FOR, identification of non-convexities) compared to NLP and quadratic constrained LP for an adopted Cigré medium voltage (MV) test system.

This paper continues research concerning an application of stochastic and metaheuristic aggregation methods in sampling the FOR using the example of the PSO. The focus is on the determination of the FOR cost structure within a hierarchical flexibility market [18], [37], [40]. The main contribution is the consideration of multiple FPUs that are aggregated by the Minkowski sum at a single bus with individual active and reactive power cost bids. First, the flexibility potentials of FPUs and the general process of the FOR determination by the PSO are described in section III. The method is introduced for non-convex FPU PQ-polygons and for the example of a two-dimensional FOR. The process can be simply adapted for the determination of a three-dimensional PQV-FOR due to slack voltage variations. The extensions regarding the FOR monetarization are described in section IV as a mixed integer LP (MILP), to specify the cost-optimal distribution of a flexibility demand to the individual FPUs per bus. The MILP can be reduced to LP in case of only convex FPU flexibility polygons. A decentral active and reactive power flexibility market based on commodity and service prices for FPU flexibility provision are introduced for the monetarization of the FOR characteristics. The presented process can be simply adapted for other population based aggregation methods like RS or further metaheuristics. The case studies in section V are based on MathWorks MATLAB simulations using the adopted Cigré MV test system [41]. The dataset of the grid model including the FPU PQ-polygons is accessible at [42] for the reproducibility of the results. Investigation aspects are the possibility in identifying zones with uniform costs within the cost structure as well as the computation time.

## III. General Process for the FOR Determination by the Particle Swarm Optimization

In the following, the general process of the FOR determination (see Fig. 1) using a PSO-based aggregation method is introduced before considering costs for a flexibility provision by the FPUs in section IV.

### *A. Description of Flexibility Potentials as PQ-Polygons*

The specific characteristics of a FPU are relevant for the FOR determination and must be communicated by the stakeholder to the corresponding system operator of a grid area (see I. in Fig. 1). The flexibility potentials of the FPU are nowadays limited by the regulatory framework of the technical guidelines. An example for the active and reactive power flexibility potential ($\Delta P_{\min}$, $\Delta P_{\max}$, $\Delta Q_{\min}$, $\Delta Q_{\max}$) of a throttled wind turbine based on the German technical guideline VDE-AR-4110 is shown in Fig. 4a) [43]. The flexibility potentials of the wind turbine are described as a polygon within the PQ-plane. In general the wind turbine manufacturers are guaranteeing a larger PQ-polygon than required [44], [45].

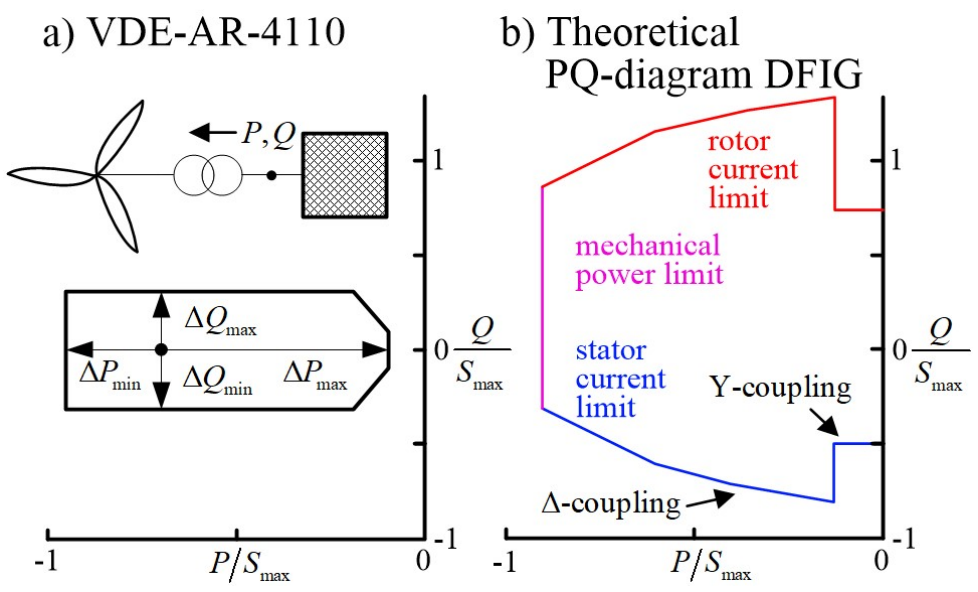

Figure 4. a) Fexibility potential of a throttled wind turbine based on the German technical guideline VDE-AR-4110 described as PQ-polygon, b) Theoretical PQ-diagram of a wind turbine (DFIG)

The flexibility potential of a FPU depends on the individual technology of the system element. The theoretical active and reactive power flexibility potential of a wind turbine with a DFIG is illustrated in Fig. 4b) [46], [47]. Compared to Fig. 4b) the technical guidelines (see Fig. 4a) lead to a loss of flexibility potentials for the system operator, which could be required and may be provided as ancillary service potential in specific system states. Differences between the technical requirements and the theoretical flexibility limits exist for each type of FPU. Fig. 5 shows six characteristic types for FPU PQ-polygons (cf. [2], [26], [30]) approximating their theoretical limits. These PQ-polygons were already used within case-studies in the context of FOR determination.

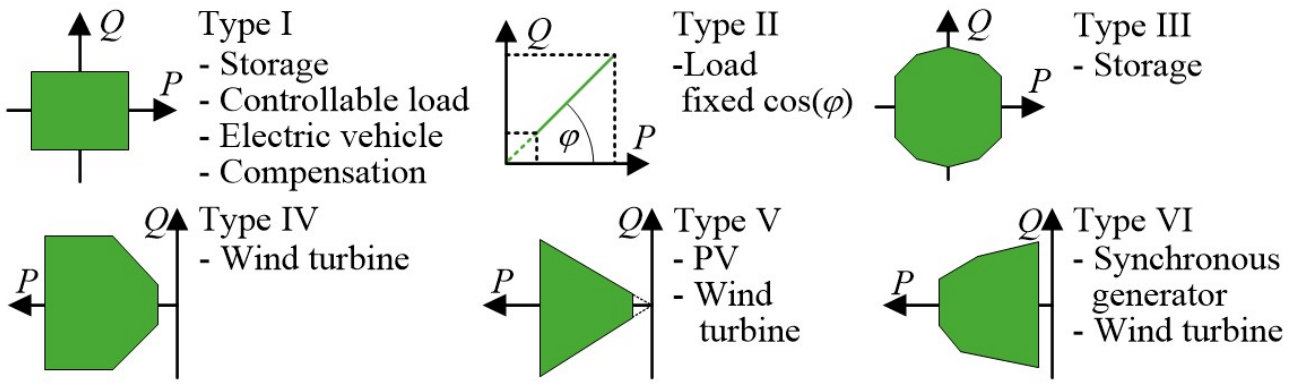

Figure 5. Technical flexibility potentials of different types of FPU described as convex PQ-polygon

Even though the PQ-polygons in Fig. 5 represent the theoretical flexibility limits of the FPUs more than the requirements within the technical guidelines there are still deviations (cf. Fig. 4b) and type VI in Fig. 5). The first reason for this is the approximation of the unit circle for the apparent power through linearization. The second reason is the convexification of the PQ-polygons for a simple description of the flexibility potentials (see [2], [48]). In general, FPU PQ-polygons can be non-convex (see Fig. 4b) [48]. Therefore, methods for the FOR determination need to consider also non-convex PQ-polygons to maximize the utilization of the FPU flexibility potentials. To guarantee this for the methods developed within this paper, the non-convex PQ-polygon of Fig. 4b) is used for wind turbines with a DFIG in the case study in section V [46].

The general FPU PQ-polygons need to be adapted prior to the FOR determination according to the specific technology (e.g. $P = 0$ for compensation of Type I, minimum power supply at Type IV, on/off-states of a FPU) [2], [33], [46]. The flexibility potentials of the FPUs are influenced by the current operating point of the system element (cf. [35]). For example, a throttled operation is necessary an increase of the active power supply of a wind turbine (see Fig. 4a). Further examples are the available primary power supply of a wind turbine or a PV-unit, the state of charge of a storage or incremental steps for the partial operation of a synchronous generator or load. For more realistic investigations, the wind forecast and the economic interests of the stakeholders can considered [35]. Nevertheless, a convex or non-convex FPU PQ-polygon results.

On load tap changing (OLTC) transformers represent another type of FPU, which can be used by the corresponding system operator for in-phase and/or quadrature voltage control [49]. OLTC transformers are not described by a PQ-polygon but instead by lower and upper boundaries and the incremental change of the voltage magnitude and angle per tap set.

The flexibility potentials of lower-level systems can be also described by a PQ-polygon representing the FOR (see I. in Fig. 1) [3], [16], [18], [50]. Thereby, the active and reactive power flexibilities of lower-level systems can be implemented analogously to PQ-polygons of FPUs by the higher-level system operator. Due to the non-linear system behavior of electric energy systems FOR PQ-polygons are in general non-convex (see Fig. 3) [9], [23], [25], [40].

## B. Sampling the FOR by the Particle Swarm Optimization

The sampling of the FOR edges for a specific voltage $V$ (see Fig. 1) is based on multiple solutions of the non-linear OPF problem defined by the objective function in Eq. (1) due to variations of $\varphi$ as well as the plus and minus signs [9], [23], [29]:

$$\min\left(\pm P_{\text{vert}}(V) \pm \tan(\varphi) Q_{\text{vert}}(V)\right) \tag{1}$$

By dividing the unit circle in $\Delta\varphi$ steps $l = 360°/\Delta\varphi$ sampling processes results (see Fig. 6).

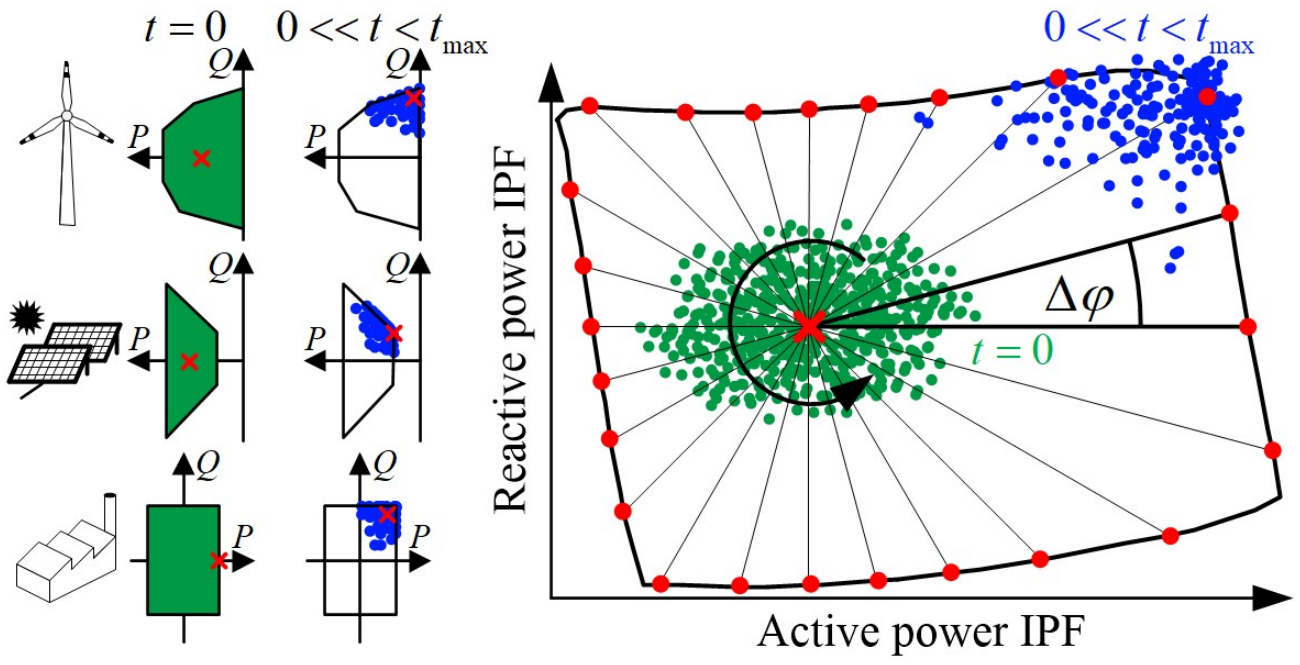

Figure 6. Schematic process of an angle-based sampling strategy for the FOR determination by a PSO-based aggregation method (swarm particles outside the FOR are violating technical constraints)

Security constraints are represented by the minimum and maximum voltage limits, the maximum thermal current of the lines as well as the rated loading of the transformers. The operational degrees of freedom are represented by the PQ-polygons of the FPUs. The population of the PSO is represented by a swarm consisting of $n$ swarm particles [39], [51], [52]. The objective of the swarm is to find the minimum of the objective function through interactions of the swarm particles during their movement through the $\mathbb{R}^{2k}$ search space, where $k$ is the number of buses. For each sampling process an individual PSO run is started. Thereby, each particle is described by a position and a velocity information. The position $\boldsymbol{x}$ of a particle represents the individual active and reactive power flexibility provision per bus by the FPUs:

$$\boldsymbol{x} = \left[\Delta P_1, \ldots, \Delta P_k, \Delta Q_1, \ldots, \Delta Q_k\right] \tag{2}$$

The velocity of a particle describes the change of the active and reactive power supplies of the FPUs for the next iteration step. At the beginning of the iterative solution process ($t = 0$) the swarm is initiated uniformly distributed within the limits of the FPUs PQ-polygons (see Fig. 6). Therefore, the FPU PQ-polygons are Delaunay triangulated by $t$ triangles and described by the vertices vectors $\Delta\boldsymbol{p}_{\text{P1}} - \Delta\boldsymbol{p}_{\text{P3}}$ and $\Delta\boldsymbol{q}_{\text{P1}} - \Delta\boldsymbol{q}_{\text{P3}}$:

$$\begin{aligned}
\Delta\boldsymbol{p}_{\text{P1}} &= [\Delta P_{\text{P1},1}, \ldots, \Delta P_{\text{P1},t}]^{\text{T}}, \Delta\boldsymbol{q}_{\text{P1}} = [\Delta Q_{\text{P1},1}, \ldots, \Delta Q_{\text{P1},t}]^{\text{T}} \\
\Delta\boldsymbol{p}_{\text{P2}} &= [\Delta P_{\text{P2},1}, \ldots, \Delta P_{\text{P2},t}]^{\text{T}}, \Delta\boldsymbol{q}_{\text{P2}} = [\Delta Q_{\text{P2},1}, \ldots, \Delta Q_{\text{P2},t}]^{\text{T}} \\
\Delta\boldsymbol{p}_{\text{P3}} &= [\Delta P_{\text{P3},1}, \ldots, \Delta P_{\text{P3},t}]^{\text{T}}, \Delta\boldsymbol{q}_{\text{P3}} = [\Delta Q_{\text{P3},1}, \ldots, \Delta Q_{\text{P3},t}]^{\text{T}}
\end{aligned} \tag{3}$$

The cumulative sum $a_{\text{r},i}$ of the triangle areas 1 to $i$ related to the complete area of the PQ-polygon is given by Eq. (4).

$$\boldsymbol{a}_{\mathrm{r,abs}} = \frac{1}{2}\left(\left(\Delta\boldsymbol{p}_{\mathrm{P1}}-\Delta\boldsymbol{p}_{\mathrm{P3}}\right)\times\left(\Delta\boldsymbol{q}_{\mathrm{P2}}-\Delta\boldsymbol{q}_{\mathrm{P3}}\right)-\left(\Delta\boldsymbol{p}_{\mathrm{P2}}-\Delta\boldsymbol{p}_{\mathrm{P3}}\right)\times\left(\Delta\boldsymbol{q}_{\mathrm{P1}}-\Delta\boldsymbol{q}_{\mathrm{P3}}\right)\right)$$
$$a_{\mathrm{r},i} = \sum\nolimits_{n=1}^{i} a_{\mathrm{r,abs},n} \Big/ \sum\nolimits_{i=1}^{t} a_{\mathrm{r,abs},i} \text{ and } \boldsymbol{a}_{\mathrm{r}} = [a_{\mathrm{r},0}=0, a_{\mathrm{r},1},\ldots,a_{\mathrm{r},t}]^{\mathrm{T}} \quad (4)$$

Uniformly distributed, random numbers $\boldsymbol{r}=[r_1,\ldots,r_j,\ldots,r_k]$ in the interval of $[0,1]$ are generated according to the swarm size $k$ to identify a specific triangle for the initialization of the swarm particle positions. A specific triangle $w$ is identified for each random number by:

$$\boldsymbol{a}_{\mathrm{r}} = [a_{\mathrm{r},0}=0, a_{\mathrm{r},1},\ldots,a_{\mathrm{r},w\text{-}1} \le r_i \le a_{\mathrm{r},w} \ldots, a_{\mathrm{r},t}]^{\mathrm{T}} \quad (5)$$

Two uniformly distributed random numbers $r_1$ and $r_2$ ($r_1 \le r_2$) within the interval of $[0,1]$ (cf. [53]) are used to determine a random point within triangle $w$:

$$\begin{aligned}\Delta P &= r_1\Delta P_{\mathrm{P1},w} + \left(r_2-r_1\right)\Delta P_{\mathrm{P2},w} + \left(1-r_2\right)\Delta P_{\mathrm{P3},w}\\ \Delta Q &= r_1\Delta Q_{\mathrm{P1},w} + \left(r_2-r_1\right)\Delta Q_{\mathrm{P2},w} + \left(1-r_2\right)\Delta Q_{\mathrm{P3},w}\end{aligned} \quad (6)$$

During the convergence process of the PSO with a maximum number of $t_{\mathrm{max}}$ iteration steps $m = t_{\mathrm{max}} n$ power flow calculations based on the Newton Raphson algorithm considering the current swarm positions are performed to evaluate the objective function value. To avoid a swarm movement outside the FPU PQ-polygons a set-to-limit operator [54] based on the point-in-polygon test according to Jordan is used. Points outside the PQ-polygon are either set to a vertex or to the nearest point on the edges by orthogonal projection. Velocities, which lead to a movement outside the PQ-polygons are inverted for an improved convergence behavior. Particles that are not complying the technical constraints are punished and represent an invalid solution. For more details regarding the PSO algorithm (e.g. general algorithm, punishment function) used in this paper see [23], [49]. The positions of the swarm particles and additional data from the power flow calculations (e.g. flexibility provision per FPU, bus voltages) represent metadata.

### C. Aggregation of FPU Flexibilities by the Minkowski Sum

In general, the performance and the quality of the results of optimization methods scale with the number of operational degrees of freedom and constraints [30], [39], [55], [56]. Furthermore, the consideration of multiple flexibilities contributing equally to the system (e.g. several FPUs at a single bus) can lead to a bad convergence behavior [48]. To avoid this, the complexity of the optimization problem is reduced by the aggregation (index: a) of the flexibility polygons $F_{\mathrm{FPU},i}$ of the $n$ FPUs connected to a single bus (see Fig. 7a).

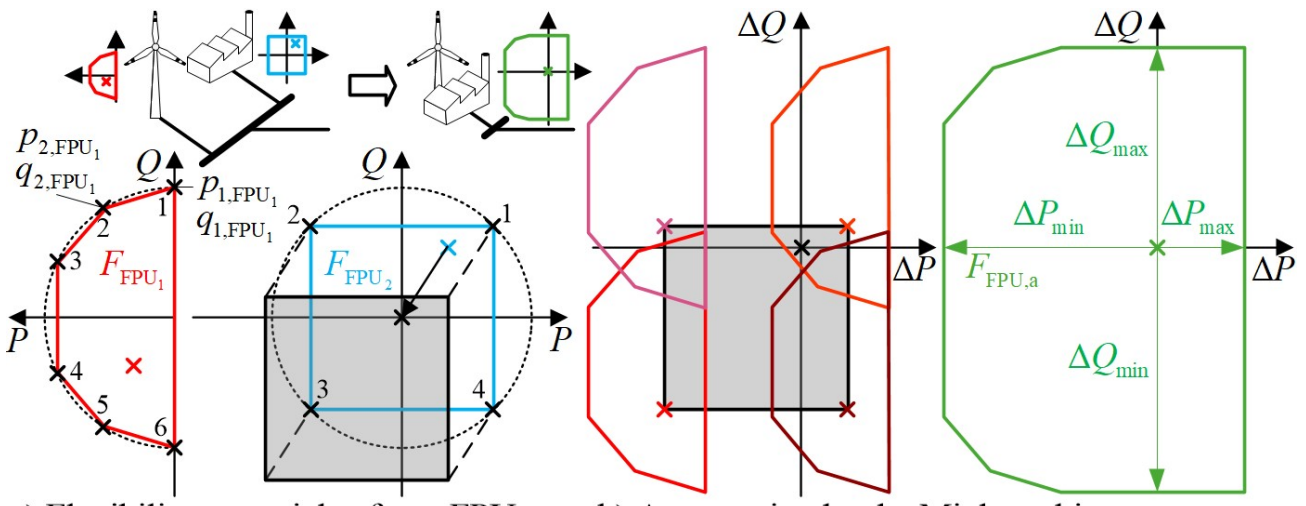

a) Flexibility potentials of two FPU b) Aggregation by the Minkowski sum

Figure 7. Schematic aggregation of two FPU flexibility polygons

A PQ-polygon $F_{\mathrm{FPU}_i}$ is represented by the coordinate vectors of the edges $\boldsymbol{p}_{\mathrm{FPU}_i}$ and $\boldsymbol{q}_{\mathrm{FPU}_i}$:

$$\boldsymbol{p}_{\mathrm{FPU}_i} = [p_{1,\mathrm{FPU}_i}, p_{2,\mathrm{FPU}_i},\ldots],\ \boldsymbol{q}_{\mathrm{FPU}_i} = [q_{1,\mathrm{FPU}_i}, q_{2,\mathrm{FPU}_i},\ldots] \quad (7)$$

Prior to the aggregation, the flexibility polygons are moved to [0,0] (see $F_{\mathrm{FPU}_2}$ in Fig. 7a) according to the current operating points (index: op). Therefore, the active and reactive power flexibilities of a FPU ($\Delta P_{\mathrm{min}}$, $\Delta P_{\mathrm{max}}$, $\Delta Q_{\mathrm{min}}$, $\Delta Q_{\mathrm{max}}$) are directly accessible:

$$\begin{aligned}&\Delta\boldsymbol{p}_{\mathrm{FPU}_i} = \boldsymbol{p}_{\mathrm{FPU}_i} - p_{\mathrm{op,FPU}_i},\ \Delta\boldsymbol{q}_{\mathrm{FPU}_i} = \boldsymbol{q}_{\mathrm{FPU}_i} - q_{\mathrm{op,FPU}_i}\\ &p_{\mathrm{op,FPU}_i\mathrm{,new}} = 0,\ q_{\mathrm{op,FPU}_i\mathrm{,new}} = 0\end{aligned} \quad (8)$$

For two convex flexibility polygons $F_{\mathrm{FPU}_1}$ and $F_{\mathrm{FPU}_2}$ the aggregated flexibility polygon $F_{\mathrm{FPU,a}}$ is determined by the Minkowski sum (see Fig. 7b), which is the totality of all sums of the edges of the individual polygons [48]:

$$F_{\mathrm{FPU,a}} = \bigcup_{i=1}^{n} F_{\mathrm{FPU}_i} = \left\{ \left(\Delta\boldsymbol{p}_{\mathrm{FPU,a}}, \Delta\boldsymbol{q}_{\mathrm{FPU,a}}\right) \left| \begin{array}{l} \Delta\boldsymbol{p}_{\mathrm{FPU,a}} = \sum_{i=1}^{n} \Delta\boldsymbol{p}_{\mathrm{FPU}_i} \\ \Delta\boldsymbol{q}_{\mathrm{FPU,a}} = \sum_{i=1}^{n} \Delta\boldsymbol{q}_{\mathrm{FPU}_i} \\ \left(\Delta\boldsymbol{p}_{\mathrm{FPU}_i}, \Delta\boldsymbol{q}_{\mathrm{FPU}_i}\right) \in F_{\mathrm{FPU}_i} \forall i \end{array} \right. \right\} \quad (9)$$

Eq. (9) is only applicable if one of the flexibility polygons is convex. If both flexibility polygons are non-convex the area can be divided into convex sub-polygons (e.g. by Delaunay triangulation) [48], [57].

## IV. Monetarization of the FOR with Cost-Optimal Flexibility Disaggregation

The monetarization of the FOR by the DSO within a hierarchical multi-level flexibility market depends on the monetarization of the FPU flexibility polygons and therefore on the assumed local active and reactive power flexibility market design [58], [59]. Different approaches for local flexibility markets are discussed in literature (cf. [37], [58], [60]). The focus of this paper is not the definition of a novel concept or the selection of an existing market concept. In contrast, requirements for a local flexibility market concept are specified resulting from the FPU monetarization approach in subsection *B*. The local flexibility market refers to [47] and to the German secondary control power market, considering the individual economic interests of the FPU stakeholders.

### A. Assumptions for the Local Flexibility Market

Today, the active power prize within the day-ahead active power market is specified by a Market Clearing Price (MCP) settlement mechanism. The MCP corresponds to the maximum selling bid price. The provision of reactive power is mandatory according to the technical guidelines and is not remunerated [47]. Without an incentive regulation, stakeholders of generating FPUs are only interested in a maximum active power supply while stakeholders of loads are interested in an active and reactive power consumption according to demand. Stakeholder of storages are interested in low market prices for loading and high market prices for unloading.

In future an availability and a utilization payment are necessary for the provision of ancillary services [47]. The availability payment (commodity prize in €/MW ) arise from guaranteed accessible flexibility potentials. An example is to fund the throttled operation of wind turbines for positive secondary control power reserve. The utilization payment (service prize in €/MWh ) results from the active and reactive power flexibility provision in a specific operating point. A local flexibility market is characterized by individual, free bids of the stakeholders for commodity and separated service prices for the active and reactive power provision. Stakeholders of FPUs with short term preserved flexibility potentials can also attend at the flexibility market with the utilization payment mechanism. The availability payment mechanism is considered prior to the FOR determination to specify a pool of FPUs for guaranteed flexibility potentials. For a specific flexibility demand (e.g. for ancillary service provision) of the higher-level system operator within the limits of the FOR the service costs are relevant. Only the utilization payment mechanism ( €/MWh ) is considered hereafter for the monetarization of the FOR.

### B. *Service Prices for the Active and Reactive Power Flexibility Provision of FPUs*

A general concept for the determination of the reactive power flexibility service costs of a FPU is based on the expected payment function (EPF) in Fig. 8 [47], [61].

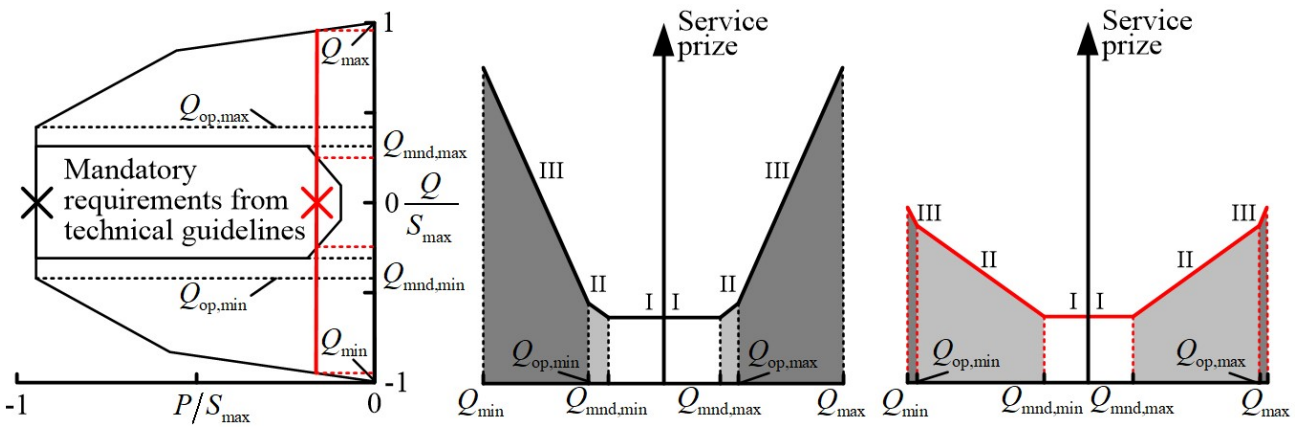

Figure 8. Schematic expected cost function of a wind turbine for two exemplary operating points [47]

Three prize zones based on individual and free bids by the stakeholder for linear cost factors (see I, II, III in Fig. 8) are used to consider the general economic interests of a specific FPU stakeholder.

- Zone I: Fixed prize to meet the limits of the mandatory requirements from the technical guidelines.
- Zone II: Variable prize for the reactive power supply for example according to quadratic increasing converter losses (in Fig. 8: linear approximation)
- Zone III: Variable prize for the reactive power supply according to a necessary active power reduction

Different price zones within the PQ-polygon of FPU (e.g. zone I) are neglected for simplification reasons but an extension of the presented approach is possible. Because of the focus of the EPF to consider only reactive power flexibilities it cannot be applied without adaptations for the determination of a FPU service prize, considering active and reactive power flexibilities. Separated service prizes for the provision of active and reactive power flexibilities as combination of zone I and II are introduced. Therefore, a linear monetarization of the active and reactive power flexibility provision $\Delta P$ and $\Delta Q$ for a duration $d$ by the cost factors $c_{\mathrm{s,P}}$ and $c_{\mathrm{s,Q}}$ is defined:

$$c_{\mathrm{s}} = \left(\Delta P \cdot c_{\mathrm{s,P}} + \Delta Q \cdot c_{\mathrm{s,Q}}\right) d \tag{10}$$

Depending on the specific FPU type Eq. (10) can be customized and different price zones of the FPU flexibility polygon may result. One example is the consideration of different costs for positive (+) and negative (-) active and reactive power changes:

$$c_{\mathrm{s}} = \left(\Delta P_{+} \cdot c_{\mathrm{s,P,+}} + \Delta P_{-} \cdot c_{\mathrm{s,P,-}} + \Delta Q_{+} \cdot c_{\mathrm{s,Q,+}} + \Delta Q_{-} \cdot c_{\mathrm{s,Q,-}}\right) d \tag{11}$$

In the following, the service costs for the flexibility provision of the FPUs is based on Eq. (10) for an appropriate but simple description of the active and reactive power flexibility service costs.

### C. *Service Prize Determination for each Swarm Particle*

The determination of the service prize $c_{\mathrm{s},\beta}$ of a single FPU connected to a bus $\beta$ is already given by Eq. (10). In this case, the metadata of a swarm particle $j$ for the flexibility provision per bus ( $\Delta P_\beta, \Delta Q_\beta$ ) is used to determine the total service prize $c_{\mathrm{s,tot},j}$ for a specific vertical active and reactive power flow $P_{\mathrm{vert}}$ and $Q_{\mathrm{vert}}$ :

$$c_{\mathrm{s,tot},j} = \sum_{\beta=1}^{k} c_{\mathrm{s},\beta} \tag{12}$$

In general, multiple FPUs with individual active and reactive power service costs and flexibility potentials are connected to a common bus $\beta$. Aggregated FPU PQ-polygons represent the summarized flexibility potentials (see section III *C*) to reduce the complexity of the optimization problem. The cost-structure of this aggregated FPU is unknown and Eq. (12) is not applicable. To enable the use of Eq. (12), the cost-structure of an aggregated FPU needs to be determined before the monetarization of the FOR. Within a premonetarization step the lower-level system operator disaggregates specific active and reactive power flexibility demands per bus ( $\Delta P_\beta, \Delta Q_\beta$ ) cost-optimal to the individual FPUs to identify the cost-structures of the aggregated FPUs. For the specification of $\Delta P_\beta$ and $\Delta Q_\beta$ a scatter within the limits of the aggregated FPU PQ-polygon is used as input for the MILP disaggregation described in the following.

### D. *Determination of the Cost-Structure of Aggregated FPUs based on MILP Disaggregation*

The disaggregation of an active and reactive power flexibility supply to the single FPU within an aggregated FPU at bus $\beta$ considering individual active and reactive power service costs represents a MILP in the general form of Eq. (13).

$$
\begin{aligned}
&\min\left( c_{\mathrm{s},\beta} = \sum_{w=1}^{2(f+g)} \lambda_w \right) \\
&\text{s.t.} \\
&\boldsymbol{A}\cdot\boldsymbol{x} \le \boldsymbol{b} \\
&\boldsymbol{A}_{\mathrm{eq}}\cdot\boldsymbol{x} = \boldsymbol{b}_{\mathrm{eq}} \\
&\boldsymbol{x}_{\min} \le \boldsymbol{x} \le \boldsymbol{x}_{\max} \\
&\boldsymbol{x} = \Big( \left[\Delta P_{\mathrm{nc},1}, \Delta Q_{\mathrm{nc},1}\right] \cdots \left[\Delta P_{\mathrm{nc},f}, \Delta Q_{\mathrm{nc},f}\right] \\
&\qquad \left[\Delta P_{\mathrm{c},1}, \Delta Q_{\mathrm{c},1}\right] \cdots \left[\Delta P_{\mathrm{nc},g}, \Delta Q_{\mathrm{nc},g}\right] \\
&\qquad \sigma_1 \cdots \sigma_f \;\; \lambda_1 \cdots \lambda_{1,2(f+g)} \Big)^{\mathrm{T}} \text{ with } \sigma_1,\ldots,\sigma_f \in \mathbb{Z}
\end{aligned}
\tag{13}
$$

Each non-convex PQ-polygon is divided into a number of convex sub-polygons (e.g. Delaunay triangulation). Enhanced approaches can be used to reduce the number of convex sub-polygons and by this the complexity of the optimization problem. The variables $\left[\Delta P_{\mathrm{nc},\mu}, \Delta Q_{\mathrm{nc},\mu}\right]$ describe a specific operating point within a FPU flexibility sub-polygon $\mu = 1,\ldots,f$. Analogously, the variables $\left[\Delta P_{\mathrm{c},\eta}, \Delta Q_{\mathrm{c},\eta}\right]$ describe a specific operating point within the initial convex FPU flexibility polygon $\eta = 1,\ldots,g$. The Boolean variable $\sigma$ represents an either-or-condition to guarantee the utilization of a single sub-polygon per non-convex FPU flexibility polygon. The variable $\lambda$ is used to determine the absolute value of the service costs within the objective function. In Eq. (14) the inequality constraints of the MILP are presented:

$$
\boldsymbol{A} = \begin{pmatrix}
\boldsymbol{C}_{\mathrm{nc}} & \boldsymbol{0} & -\boldsymbol{b}_{\mathrm{nc}} & \boldsymbol{0} & \boldsymbol{0} \\
\boldsymbol{0} & \boldsymbol{C}_{\mathrm{c}} & \boldsymbol{0} & \boldsymbol{0} & \boldsymbol{0} \\
\boldsymbol{I}_{\mathrm{nc}} & \boldsymbol{0} & -\boldsymbol{M}_{\max} & \boldsymbol{0} & \boldsymbol{0} \\
-\boldsymbol{I}_{\mathrm{nc}} & \boldsymbol{0} & \boldsymbol{M}_{\min} & \boldsymbol{0} & \boldsymbol{0} \\
\boldsymbol{Co}_{\mathrm{nc}} & \boldsymbol{0} & \boldsymbol{0} & -\boldsymbol{I}_{\mathrm{nc}} & \boldsymbol{0} \\
\boldsymbol{0} & \boldsymbol{Co}_{\mathrm{c}} & \boldsymbol{0} & \boldsymbol{0} & -\boldsymbol{I}_{\mathrm{c}} \\
-\boldsymbol{Co}_{\mathrm{nc}} & \boldsymbol{0} & \boldsymbol{0} & -\boldsymbol{I}_{\mathrm{nc}} & \boldsymbol{0} \\
\boldsymbol{0} & -\boldsymbol{Co}_{\mathrm{c}} & \boldsymbol{0} & \boldsymbol{0} & -\boldsymbol{I}_{\mathrm{c}}
\end{pmatrix},
\boldsymbol{b} = \begin{pmatrix} \boldsymbol{0} \\ \boldsymbol{b}_{\mathrm{c}} \\ \boldsymbol{0} \\ \boldsymbol{0} \\ \boldsymbol{0} \end{pmatrix}
\tag{14}
$$

The matrices $\boldsymbol{C}_{\mathrm{nc}}$, $\boldsymbol{C}_{\mathrm{c}}$, $\boldsymbol{B}_{\mathrm{nc}}$ and the vector $\boldsymbol{b}_{\mathrm{c}}$ describe the limitation of the solution space by linear equations ($m_1 \cdot x_1 + m_2 \cdot x_2 = b$) based on the individual number of edges $\kappa$ of each FPU flexibility polygon:

$$
\boldsymbol{C}_{\mathrm{nc}} = \begin{pmatrix} \boldsymbol{C}_{\mathrm{nc},1} & & 0 \\ & \ddots & \\ 0 & & \boldsymbol{C}_{\mathrm{nc},f} \end{pmatrix},\;
\boldsymbol{B}_{\mathrm{nc}} = \begin{pmatrix} \boldsymbol{b}_{\mathrm{nc},1} & & 0 \\ & \ddots & \\ 0 & & \boldsymbol{b}_{\mathrm{nc},f} \end{pmatrix}
\tag{15}
$$

$$
\begin{aligned}
&\boldsymbol{C}_{\mathrm{c}} = \begin{pmatrix} \boldsymbol{C}_{\mathrm{c},1} & & 0 \\ & \ddots & \\ 0 & & \boldsymbol{C}_{\mathrm{c},g} \end{pmatrix},\;
\boldsymbol{b}_{\mathrm{c}} = \begin{pmatrix} \boldsymbol{b}_{\mathrm{c},1} \\ \vdots \\ \boldsymbol{b}_{\mathrm{c},g} \end{pmatrix} \\
&\boldsymbol{C}_{\mathrm{nc},\mu}, \boldsymbol{C}_{\mathrm{c},\eta} = \begin{pmatrix} m_{1,1} & m_{1,2} \\ \vdots & \vdots \\ m_{\kappa,1} & m_{\kappa,2} \end{pmatrix},\;
\boldsymbol{b}_{\mathrm{nc},\mu}, \boldsymbol{b}_{\mathrm{c},\eta} = \begin{pmatrix} b_1 \\ \vdots \\ b_\kappa \end{pmatrix}
\end{aligned}
\tag{16}
$$

The matrices $\boldsymbol{I}_{\mathrm{nc}}$ and $\boldsymbol{I}_{\mathrm{c}}$ are identity matrices of the size $(f \times f)$ and $(g \times g)$, respectively. Line 3 to 4 in Eq. (14) belong to an if-then-condition to consider the $\sigma$ corresponding constraints for non-convex FPU flexibility polygons which are divided into multiple convex sub-polygons. The matrix $\boldsymbol{M}_{\max}$ include the maximum active and reactive power values of the corresponding convex sub-polygon edges $\Delta\boldsymbol{p}_\mu$ and $\Delta\boldsymbol{q}_\mu$:

$$
\boldsymbol{M}_{\max} = \begin{pmatrix}
\begin{pmatrix} \max(\Delta\boldsymbol{p}_1) \\ \max(\Delta\boldsymbol{q}_1) \end{pmatrix} & & 0 \\
& \ddots & \\
0 & & \begin{pmatrix} \max(\Delta\boldsymbol{p}_f) \\ \max(\Delta\boldsymbol{q}_f) \end{pmatrix}
\end{pmatrix}
\tag{17}
$$

For the determination of the matrix $\boldsymbol{M}_{\min}$ the Eq. (17) is adapted for the minimum active and reactive power values of the corresponding convex sub-polygon edges $\Delta\boldsymbol{p}_\mu$ and $\Delta\boldsymbol{q}_\mu$. Line 5 to 8 in Eq. (14) are used to determine the absolute value $\lambda$ of the service costs ($c_{\mathrm{s,P}}$, $c_{\mathrm{s,Q}}$) by using the cost matrices $\boldsymbol{Co}_{\mathrm{nc}}$ and $\boldsymbol{Co}_{\mathrm{c}}$, e.g.:

$$
\boldsymbol{Co}_{\mathrm{nc}} = \begin{pmatrix}
c_{\mathrm{s,P},1} & & & & \\
& c_{\mathrm{s,Q},1} & & & \\
& & \ddots & & \\
& & & c_{\mathrm{s,P},\mu} & \\
& & & & c_{\mathrm{s,Q},\mu}
\end{pmatrix}
\tag{18}
$$

The equality constraints are given by:

$$
\begin{aligned}
&\boldsymbol{A}_{\mathrm{eq}} = \begin{pmatrix}
\boldsymbol{I}_{\mathrm{nc},1} \cdots \boldsymbol{I}_{\mathrm{nc},f} & \boldsymbol{I}_{\mathrm{c},1} \cdots \boldsymbol{I}_{\mathrm{c},g} & \boldsymbol{0} \cdots \boldsymbol{0} & \boldsymbol{0} \cdots \boldsymbol{0} & \boldsymbol{0} \cdots \boldsymbol{0} \\
& & \boldsymbol{e}_1 \quad \boldsymbol{0} & & \\
\boldsymbol{0} & \boldsymbol{0} & \ddots & \boldsymbol{0} & \boldsymbol{0} \\
& & \boldsymbol{0} \quad \boldsymbol{e}_o & &
\end{pmatrix} \\
&\boldsymbol{b}_{\mathrm{eq}} = \begin{pmatrix} \Delta P_\beta & \Delta Q_\beta & 1 & \cdots & 1 \end{pmatrix}^{\mathrm{T}} \text{ with } e = (1,\ldots,1)
\end{aligned}
\tag{19}
$$

Lines 1 and 2 in Eq. (19) guarantee the compliance of the active and reactive power demand $\Delta P_\beta$ and $\Delta Q_\beta$ from the specific aggregated FPU flexibility polygon at bus $\beta$. $\boldsymbol{I}_{\mathrm{nc},\mu}$ and $\boldsymbol{I}_{\mathrm{c},\eta}$ are identity matrices of size $(2 \times 2)$. The vector $\boldsymbol{e}$ assign the convex sub-polygons to the corresponding non-convex, aggregated FPU flexibility polygon within an either-or-condition. The minimum and maximum limits for the flexibility variables are given by:

$$
\begin{aligned}
\boldsymbol{x}_{\min} = \Big( &\left[\min(\Delta\boldsymbol{p}_1), \min(\Delta\boldsymbol{q}_1)\right] \cdots \left[\min(\Delta\boldsymbol{p}_f), \min(\Delta\boldsymbol{q}_f)\right] \\
&\left[\min(\Delta\boldsymbol{p}_1), \min(\Delta\boldsymbol{q}_1)\right] \cdots \left[\min(\Delta\boldsymbol{p}_g), \min(\Delta\boldsymbol{q}_g)\right] \\
&0 \cdots 0 \quad 0 \cdots 0 \quad 0 \cdots 0 \Big)^{\mathrm{T}}
\end{aligned}
\tag{20}
$$

$$
\begin{aligned}
\boldsymbol{x}_{\max} = \Big( &\left[\max(\Delta\boldsymbol{p}_1), \min(\Delta\boldsymbol{q}_1)\right] \cdots \left[\max(\Delta\boldsymbol{p}_f), \max(\Delta\boldsymbol{q}_f)\right] \\
&\left[\max(\Delta\boldsymbol{p}_1), \min(\Delta\boldsymbol{q}_1)\right] \cdots \left[\max(\Delta\boldsymbol{p}_g), \max(\Delta\boldsymbol{q}_g)\right] \quad 0 \cdots 0 \\
&\left[c_{\mathrm{s,P,1}} \max(|\Delta\boldsymbol{p}_1|), c_{\mathrm{s,Q,1}} \max(|\Delta\boldsymbol{q}_1|)\right] \cdots \left[c_{\mathrm{s,P,f}} \max(|\Delta\boldsymbol{p}_f|), c_{\mathrm{s,Q,f}} \max(|\Delta\boldsymbol{q}_f|)\right] \\
&\left[c_{\mathrm{s,P,1}} \max(|\Delta\boldsymbol{p}_1|), c_{\mathrm{s,Q,1}} \max(|\Delta\boldsymbol{q}_1|)\right] \cdots \left[c_{\mathrm{s,P},g} \max(|\Delta\boldsymbol{p}_g|), c_{\mathrm{s,Q},g} \max(|\Delta\boldsymbol{q}_g|)\right] \Big)^{\mathrm{T}}
\end{aligned}
\tag{21}
$$

The total service prize $c_{\mathrm{s,tot},j}$ for each particle $j$ can be determined according to Eq. 12 based on the presented MILP disaggregation method. An example for the determination of the cost-structure of an aggregated FPU within the premonetarization step is presented within the case study in section V.

### *E. Determination of the FOR Cost Structure*

An overlay of the monetarized $t_{\max}$ particle swarms lead to overlapping points within the cost structure of the FOR (see Fig. 3a) and an identification of cost zones is not possible. Overlapping points represent swarm particles with similar values for $P_{\text{vert}}$ and $Q_{\text{vert}}$ but different flexibility provisions per bus $\boldsymbol{x} = [\Delta P_1, \ldots, \Delta P_k, \Delta Q_1, \ldots, \Delta Q_k]$. For the higher-level system operator only the particle with the minimum summarized service prize of the FPUs is relevant for a cost-optimal flexibility provision.

The costs for the active power losses $P_{\text{loss},j}$ of particle $j$ and a duration $d$ are determined by Eq. (22) and added to $c_{\text{s,tot},j}$:

$$c_{\text{loss},j} = \left(P_{\text{loss},j} - P_{\text{loss},0}\right) c_{\text{loss}} d \;\forall \left(P_{\text{loss},j} - P_{\text{loss},0}\right) > 0 \qquad (22)$$

Within Eq. (22) only increased grid losses compared to the initial losses $P_{\text{loss},0}$ are monetarized for the higher-level system operator. The idea behind this is that a reduction of grid losses is in the economic interest of the lower-level system operator.

For the determination of the FOR cost-structure, the swarm particles within the whole iteration process of the PSO are assigned to a scatter within the limits of the FOR PQ-polygon. The scatter is described by the active and reactive power vectors $\boldsymbol{p}_{\text{sc}}$ and $\boldsymbol{q}_{\text{sc}}$ (see Fig. 9). For the determination of the cost structure $\boldsymbol{c}_{\text{sc}}$ first, the Euclidean distance between the active and reactive power interconnection power flow of a swarm particle $(P_{\text{vert},j}, Q_{\text{vert},j})$ and the scatter points $(P_{\text{sc},y}, Q_{\text{sc},y})$ is identified (see Fig. 9a):

$$d_{j,y} = \sqrt{\left(P_{\text{sc},y} - P_{\text{vert},j}\right)^2 + \left(Q_{\text{sc},y} - Q_{\text{vert},j}\right)^2} \;\forall\; j = 1,\ldots,m \wedge y = 1,\ldots,h \qquad (23)$$

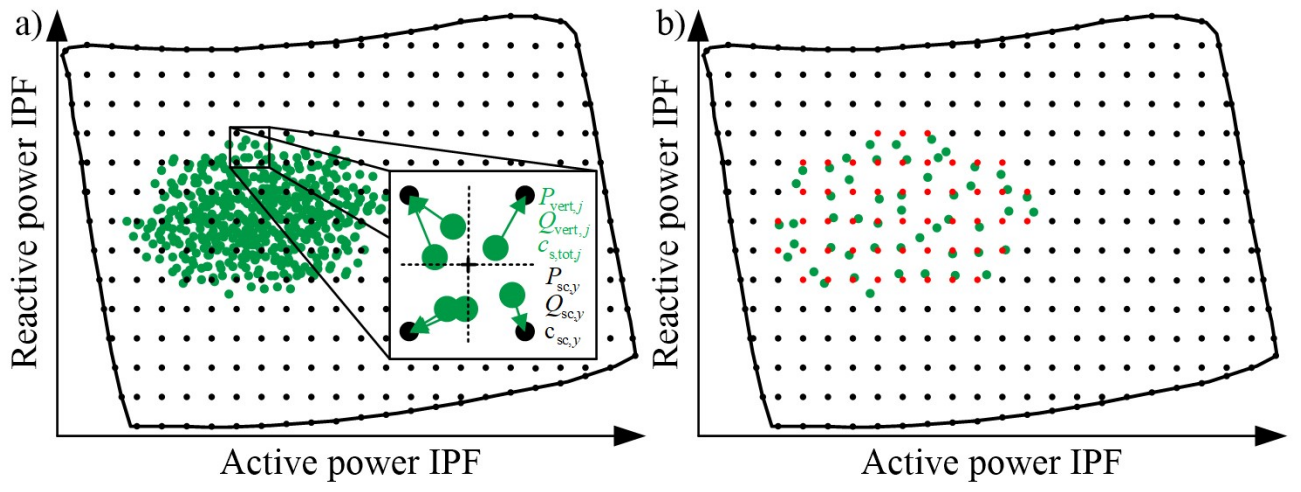

Figure 9. a) Schematic assignment of the particle swarm of a specific iteration step within the PSO to the scatter, b) particles (green) with the minimum costs for the corresponding scatter point (red)

For the scatter point $y$ with the minimum distance to the position of a swarm particle $j$ the cost value $c_{\text{sc},y}$ is updated by the particle costs $c_{\text{s,tot},j}$, if:

$$c_{\text{sc},y} \left(c_{\text{sc},y} > c_{\text{s,tot},j}\right) = c_{\text{s,tot},j} \qquad (24)$$

Based on Eq. (24) the particles with the minimum costs for the corresponding scatter points are identified (see Fig. 9b). The monetarization of the FOR is completed by the determination of the cost structure.

## V. CASE-STUDY

An adaptation of the Cigré medium voltage system is used for the case study [41]. The original grid was extended by aggregated low voltage grids and a variety of different FPUs. Each type of FPU is working at the same operating point and has the same flexibility potentials adapted from [2]. In scenario 1, the FPU PQ-polygons correspond to Fig. 10. In scenario 2 the installed power of the FPUs is halved. The high voltage bus is used as slack with a specified voltage of $V_{\text{slack}} = 110\text{ kV e}^{j0^\circ}$. Information on the lines and the transformers are given in Table I and II. The voltage change per tap-set of the OLTC HV/MV-transformer at the higher-voltage bus is 0.25% related to the corresponding rated voltage. The maximum and minimum tap-set is limited to ±10 steps. Service costs for the OLTC transformer are neglected within this paper. In general, the service prize of a OLTC transformer can be estimated by the equivalent lifetime loss per tap-set [17]. The full data set, including the FPU flexibility polygons, is available at [42].

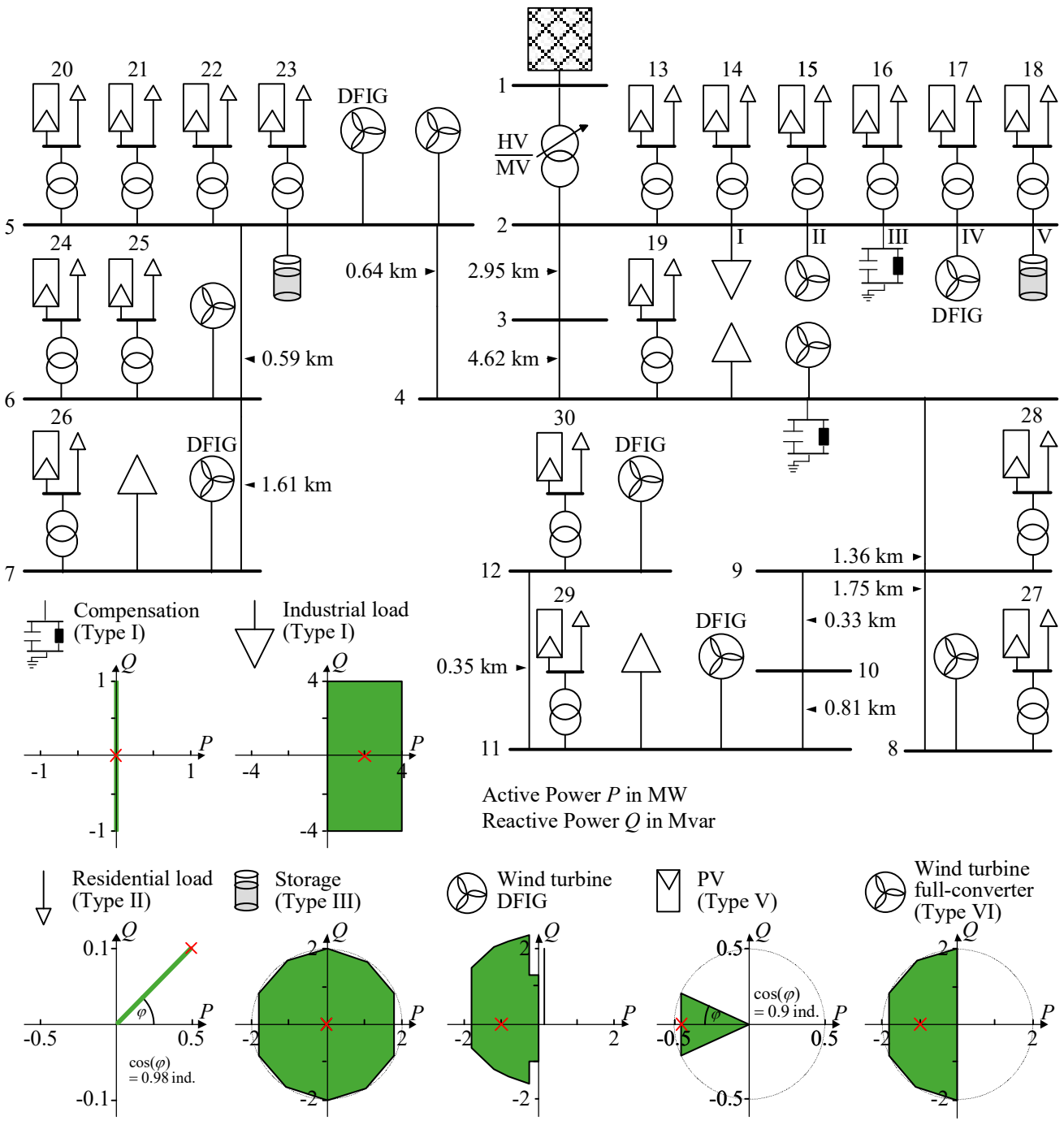

Figure 10. Topology of the adapted Cigré medium voltage test system including operating points and flexibility potentials of the FPU

TABLE I. LINE DATA

| Maximum thermal line current $I_{\text{th,max}}$ | Line constant for the loop resistance $R'$ | Line constant for the loop inductance $L'$ | Line constant for the loop capacitance $C'$ |
|---|---|---|---|
| 680 A | 0.501 Ω / km | 2.279 mH / km | 0.151 μF / km |

TABLE II. TRANSFORMER DATA

| High voltage $V_{\text{HV}}$ | Low voltage $V_{\text{LV}}$ | Rated loading $S_{\text{r}}$ | Short circuit voltage $v_{\text{sc}}$ | Copper loss $P_{\text{Cu}}$ | Open circuit current $i_{\text{oc}}$ | Iron loss $P_{\text{Fe}}$ |
|---|---|---|---|---|---|---|
| 110 kV | 20 kV | 25 MVA | 12% | 25 kW | 0.2% | 0 kW |
| 20 kV | 0.4 kV | 2 MVA | 8% | 16.7 kW | 0.2% | 4 kW |

TABLE III. COST FACTORS AT THE LOW VOLTAGE LEVEL IN €/MWh

| Bus | 13 | 14 | 15 | 16 | 17 | 18 | 19 | 20 | 21 |
|---|---|---|---|---|---|---|---|---|---|
| $c_{s,P}$ | PV: 50<br>Load: 80 | 30<br>70 | 90<br>50 | 40<br>20 | 50<br>90 | 10<br>80 | 10<br>50 | 10<br>90 | 50<br>60 |
| $c_{s,Q}$ | PV: 0.6<br>Load: 0.7 | 0.7<br>0.7 | 0.6<br>1.0 | 1.0<br>0.9 | 0.5<br>1.0 | 0.6<br>0.7 | 0.8<br>0.1 | 0.3<br>0.9 | 0.7<br>0.4 |
| Bus | 22 | 23 | 24 | 25 | 26 | 27 | 28 | 29 | 30 |
| $c_{s,P}$ | PV: 40<br>Load: 70 | 30<br>20 | 70<br>80 | 90<br>90 | 90<br>30 | 70<br>70 | 40<br>40 | 50<br>10 | 10<br>20 |
| $c_{s,Q}$ | PV: 0.3<br>Load: 0.5 | 0.7<br>0.5 | 0.4<br>0.4 | 0.9<br>0.2 | 0.3<br>0.4 | 1.0<br>0.6 | 0.9<br>0.5 | 0.4<br>0.7 | 0.9<br>0.2 |

TABLE IV. COST FACTORS AT THE MEDIUM VOLTAGE LEVEL IN €/MWh

| Bus | 2 (see case 1) | 4 | 5 | 6 | 7 | 8 | 11 | 12 |
|---|---|---|---|---|---|---|---|---|
| $c_{s,P}$ | Industrial load: 90<br>Wind turbine: 30<br>DFIG: 40<br>Storage: 60 | 60<br>30<br>-<br>- | -<br>40<br>10<br>70 | -<br>20<br>-<br>- | 80<br>-<br>40<br>- | -<br>60<br>-<br>- | 90<br>-<br>60<br>- | -<br>-<br>70<br>- |
| $c_{s,Q}$ | Industrial load: 0.9<br>Wind turbine: 0.1<br>Compensation: 0.3<br>DFIG: 0.5<br>Storage: 0.7 | 0.2<br>1.0<br>1.0<br>-<br>- | -<br>0.8<br>-<br>0.5<br>0.6 | -<br>0.5<br>-<br>-<br>- | 0.6<br>-<br>-<br>0.5<br>- | -<br>0.2<br>-<br>-<br>- | 0.7<br>-<br>-<br>0.4<br>- | -<br>-<br>-<br>0.2<br>- |

The number of samples for the FOR determination is set to 45 with a corresponding sampling angle of $\Delta\varphi = 8°$ in Eq. (1). The swarm size $n$ and the maximum iteration step $t_{\max}$ are set to 200, which leads to 1.8 million power flow calculations. For a detailed parameterization of the PSO see [23]. The sampling process is fully parallelized on 45 workers with usual computation power in MathWorks MATLAB.

The prize $c_{\text{loss}}$ in Eq. (22) is set to 50 €/MWh according to typical German market prices. The service cost factor (see Table III and IV) for the active power flexibility provision $c_{s,P}$ is determined randomly in an interval of [10, 20,…, 90] for each FPU. The service cost factor for the reactive power flexibility provision $c_{s,Q}$ is determined in an interval of [0.1, 0.2,…, 1]. The technology specific reasons for the individual service costs of a FPU are neglected for the investigations within this paper. The duration of the flexibility provision is set to $d = 1$.

The objective of the case study is the application and evaluation of the process for the determination of the FOR cost structure, presented in section IV. The investigations are divided into two steps. First, the cost structure of the aggregated FPU PQ-polygon is premonetarized by the MILP presented in section IV *D*. Exemplarily, the five FPUs at bus 2 and the FPU PQ-polygons of scenario 1 are used. Second, in both investigation scenarios, the monetarized FOR at the vertical system interface is determined based on the methods described in sections IV *C* and IV *E*.

## *A. Premonetarization of Aggregated FPU PQ-Polygon*

The composition of the PQ-polygon of the aggregated FPU at bus 2 is presented in Fig. 11a). Starting from the load (see I in Fig. 10) the flexibility potentials of the other FPUs are added by the Minkowski sum. The PQ-polygon of the aggregated FPU is non-convex due to the DFIG flexibility potentials. The MILP disaggregation (see section III *D*) is applied for a 100x100 scatter in the limits of the aggregated FPU PQ-polygon for the determination of the cost structure (see Fig. 11b). Further points are specified by the edges of each subpolygon combination during the Minkowski process in Fig. 11a).

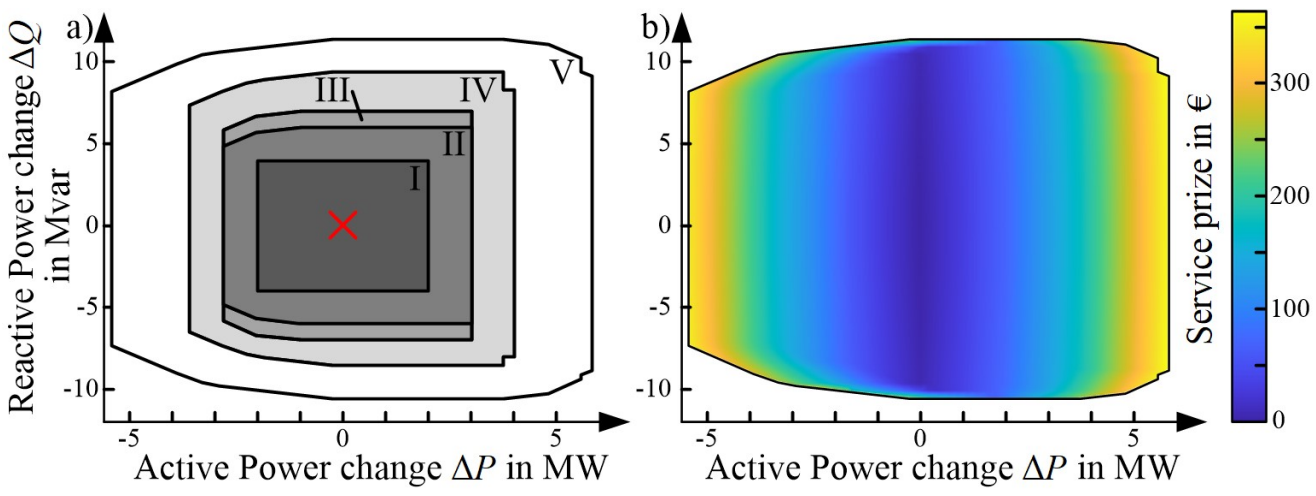

Figure 11. a) Aggregation of the active and reactive power flexibility potentials of the FPUs at bus bus 2, b) Premonetarization of the aggregated FPU by service costs based on MILP disaggregation

The linear service cost factors $c_{s,P}$ and $c_{s,Q}$ of the FPUs at bus 2 leads to a two-dimensional piecewise linear function for the cost structure of the aggregated FPU. The information of the cost structure can be used for a cost-optimal distribution of an active and reactive power flexibility demand at bus 2 on the individual FPUs. The computation time is 0.9 s.

For the premonetarization of the three FPUs at bus 5 the computation time was 0.5 s and for the busses with two FPUs about 0.2 s. The computation time depends on the complexity of the MILP and the number of MILP runs, which depends on the resolution of the meshgrid. With a higher number of FPUs within the aggregated FPU the computation time for the premonetarization will increase. To avoid this, the FPUs can be sorted and aggregated regarding equal economic interests of the stakeholders prior to the premonetarization. The computation will be also increased for solving a mixed-integer quadratic programming problem in the case of quadratic cost functions for the FPUs active and reactive power service prizes. These aspects become important at more realistic grid scenarios with a variety of different FPUs at a single bus.

## *B. Determination of the FOR Cost Structure in Scenario 1*

Within the heat map of Fig. 12a) the density of the swarm particles during the PSO-based sampling process and the resulting FOR are presented. The edges of the FOR are sampled more detailed then the center of the FOR, which results from the convergence behavior of the PSO (cf. Fig. 6). The FOR is limited by the technical constraints of the grid (see Fig. 12b) and not by the flexibility limits of the FPUs.

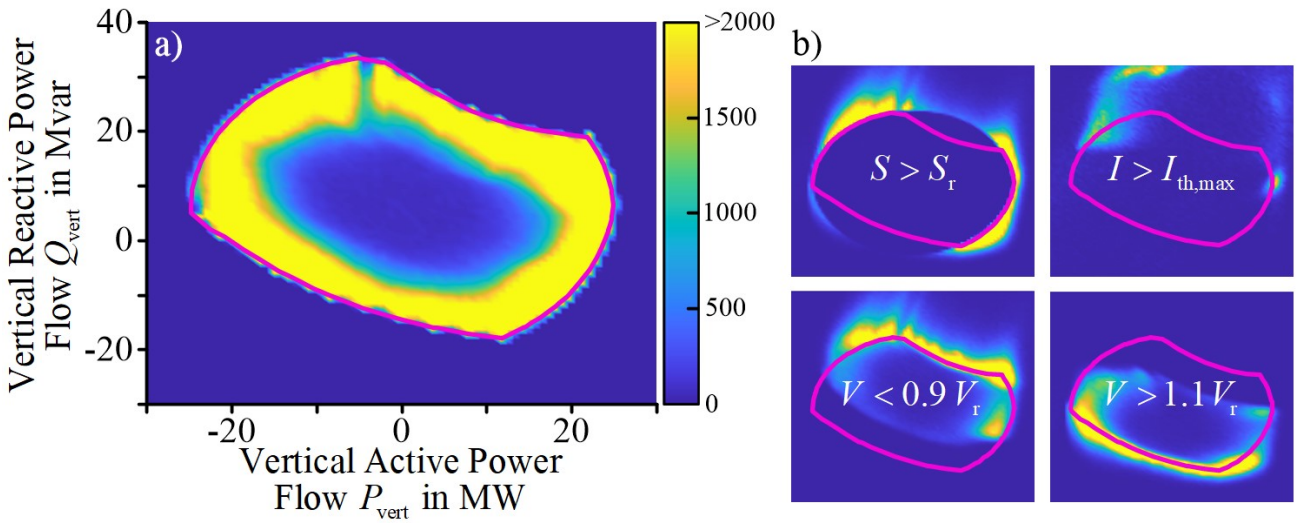

Figure 12. Scenario 1: a) Density of the swarm particles during the PSO-based sampling process, b) Limitation of the FOR by technical constraints (rated power HV/MV-transformer, maximum thermal line currents, lower and upper voltage limits)

The results of the FOR including the service-costs of the FPUs and the costs for increased grid losses within the distribution grid are presented in Fig. 13a) and b) based on a 100x100 scatter. Within the FOR the cost zones are blurred, which results from smooth cost gradients. The only exception are the areas near the maximum active power supply and reactive power consumption of the lower-level system. Considering the density of the swarm in Fig. 12a), these areas are sampled by fewer swarm particles, which also result in less metadata.

From this it can be derived that for areas with a low swarm particle density in consequence of a challenging system state (here: high voltages, high transformer loading) the cost structure becomes less uniform. Another reason is that the flexibility potentials of the FPUs are larger than the maximum power transfer capability of the HV/MV-transformer. Thereby, more constellations of the FPUs to guarantee a specific IPF are possible. The monetarization of the FOR is only based on metadata because the reduction of the service costs is not an objective within the sampling process of the PSO. This results in regions with varying service costs (see Fig. 13a) in the usually uniform cost-structure. The non-uniform transition of the cost structure in the middle of the FOR results from the lower density of the swarm particles in this area.

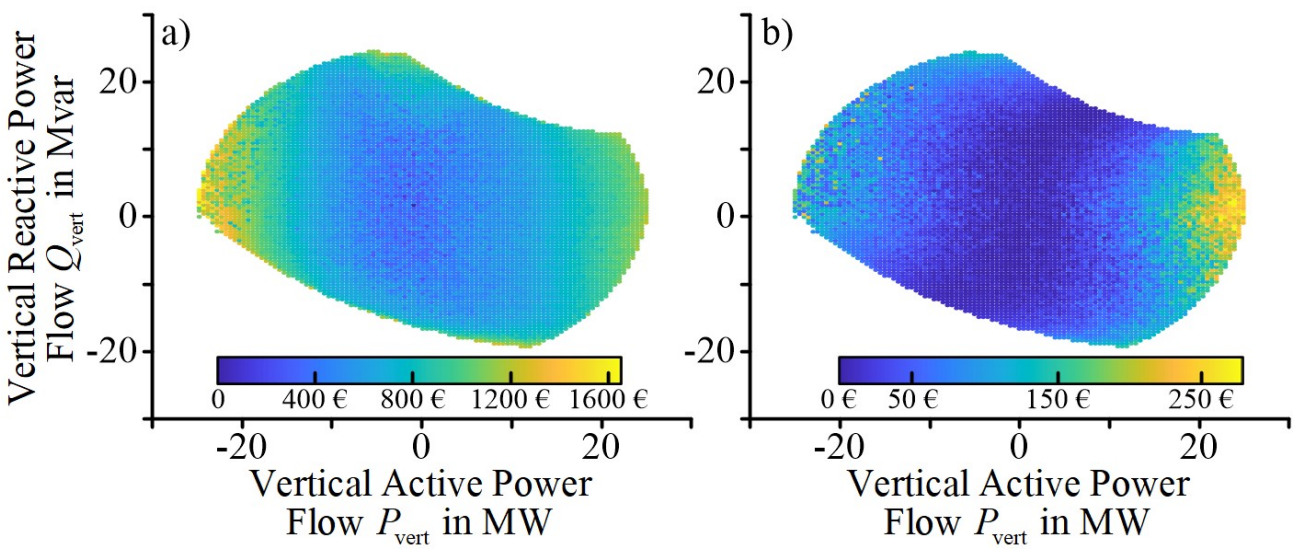

Figure 13. Scenario 1: a) FOR including service costs of the FPUs, b) FOR including costs for increased active power losses

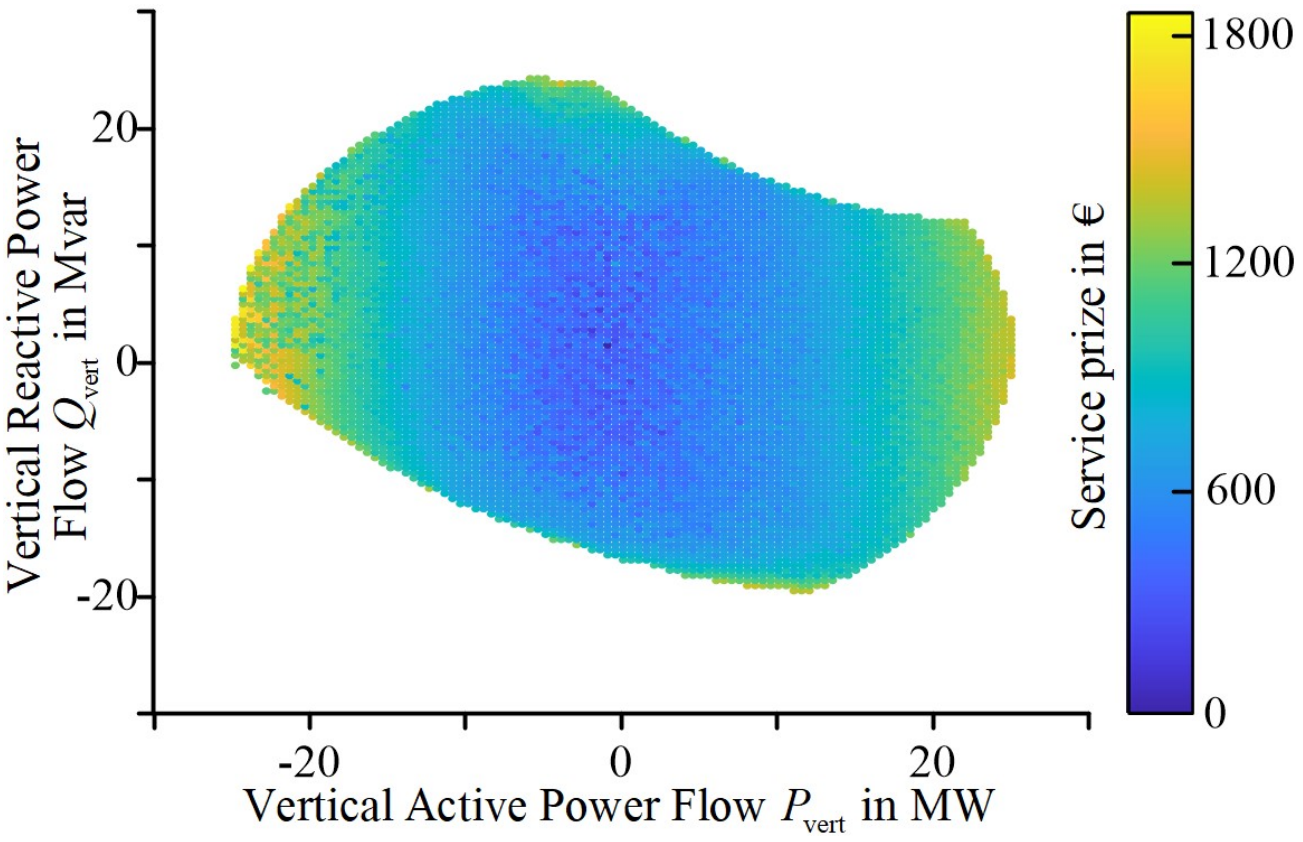

Figure 14. Complete cost-structure of the FOR in scenario 1

Within the case-study only higher active power losses compared to the initial system state are monetarized in Eq. (22). Thereby, the dark blue area in the bottom, middle and top of the FOR in Fig. 13b) result. The grid losses are increased especially in case of high active power transfers at the vertical system interconnection in case of high load or supply. The transition of the service costs within the complete cost-structure of the FOR (see Fig. 14) is similar to the results in Fig. 13a). The impact of the active power losses is only significant in the area with a high active power transfer to/from the lower-level system.

In Fig. 15a) and b) the cost structure of the FOR is presented in more detail. The point with the lowest service prizes is represented by the initial system state without any flexibility provision by the FPUs.

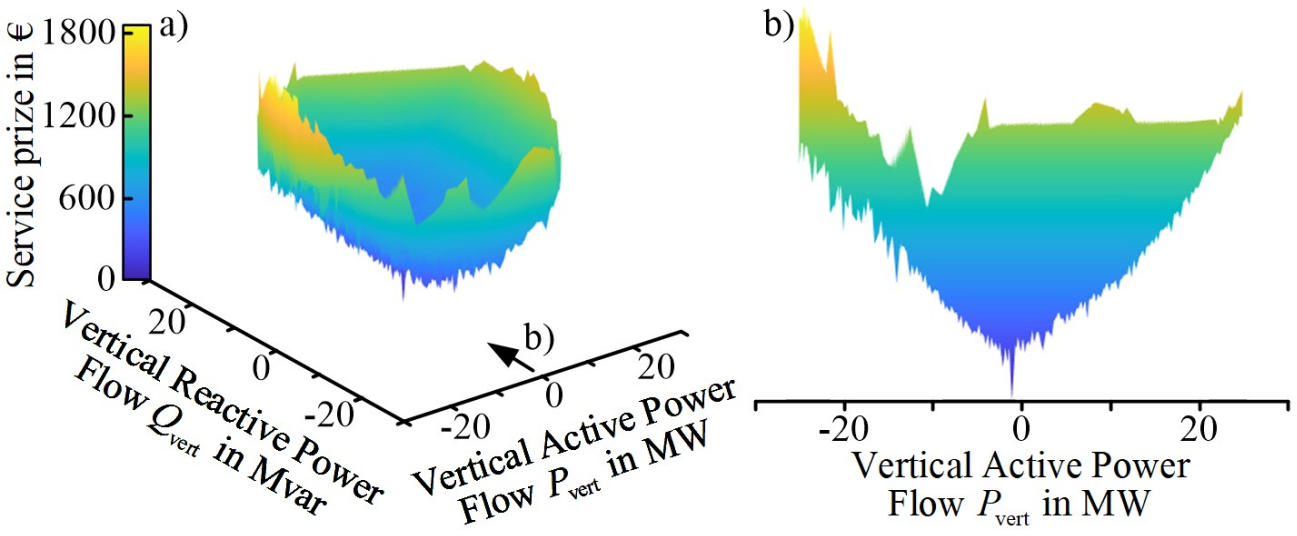

Figure 15. Scenario 1: a) Threedimensional cost-structure of the FOR, b) front view

Especially for particles located in the middle of the FOR the flexibility provision of the FPUs is not representing the minimum service costs. In general, the higher-level system operator can assume the results for the cost structure as estimation for the real costs to be excepted for a specific vertical active and reactive power flow. The computation time for the FOR sampling is 56 s and for the determination of the cost-structure 6 s.

## *C. Determination of the FOR Cost Structure in Scenario 2*

In scenario 2 the density of the swarm particles within the FOR is higher than in scenario 1 (see Fig. 16a). The top and bottom edges of the FOR are limited, analogously to scenario 1, by the lower and upper voltage limits (see Fig. 16b). The left and the right side of the FOR are limited by the flexibility limits of the FPU PQ-polygons.

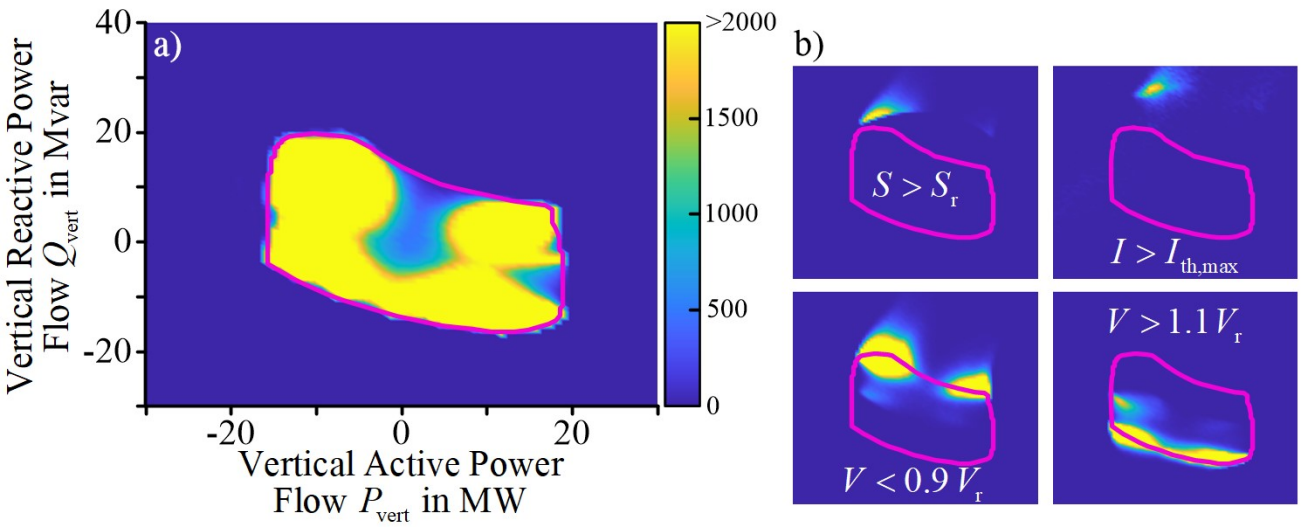

Figure 16. Scenario 2: a) Density of the swarm particles during the PSO-based sampling process, b) Limitation of the FOR by technical constraints (rated power HV/MV-transformer, maximum thermal line currents, lower and upper voltage limits)

Within the cost-structure of the FOR in Fig. 17 clear cost zones can be identified. This also applies for regions with a low density of the swarm particles like the maximum active power consumption (see Fig. 16a). The variations of the cost gradients are more significant compared to scenario 1. This can be identified by the width of the individual color ranges.

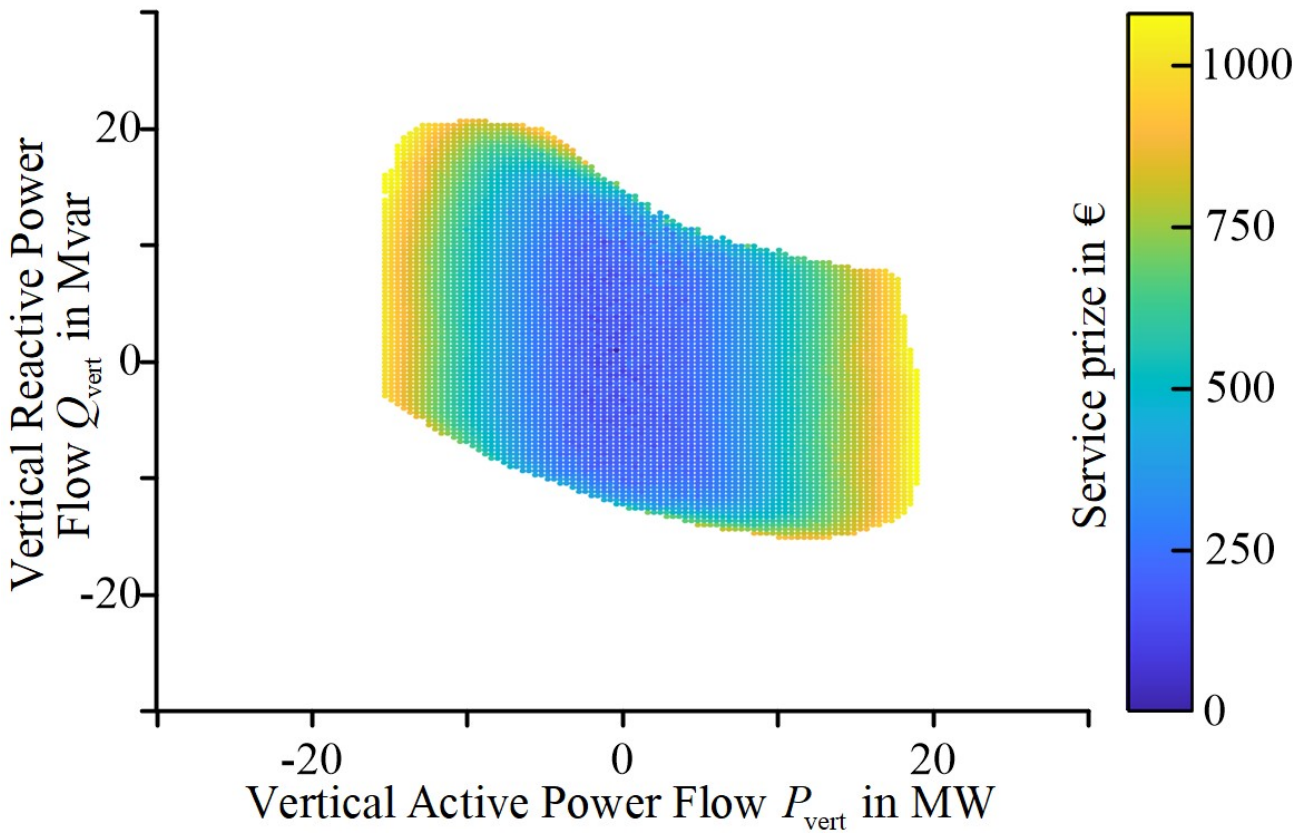

Figure 17. Complete cost-structure of the FOR in scenario 2

The computation time for the FOR sampling in scenario 2 is 58 s, which can be explained by numerical performance variations. The computation time for the determination of the cost-structure is 8 s. The increased computation time results from the higher number of swarm particles complying the technical constraints within the FOR compared to scenario 1. Within scenario 1 only 1.2 million swarm particles are included in the FOR compared to 1.6 million in scenario 2.

## VI. Discussion and Conclusions

The presented approach for the monetarization of the FOR leads to comprehensible results within the case-study of the Cigré medium voltage system and the additional computation time for processing the metadata is small. In both investigation scenarios, cost-structures with mostly clear cost zones are obtained. In general, the cost structure is blurred in regions with a low sampling density. Significantly more definable cost zones can be achieved by combining multiple costs to a common zone (e.g. 40 €-50 €, cf. [14]). The computation time for the premonetarization of the aggregated FPUs and the processing of the metadata will increase for larger systems with a variety of buses and FPUs. Investigations on larger systems are necessary to evaluate the performance of the algorithms and the quality of the results in more detail. Nevertheless, the presented approach is suitable for an application within the day-ahead and intraday operational management.

The cost-structures are representing an estimation of the service costs that can be assumed by the higher-level system operator for a specific adaptation of the IPF and the corresponding flexibility provision of the FPUs. The cost-optimal distribution of a flexibility demand (e.g. ancillary service provision) from the higher-level system operator to the FPUs within the lower-level system is guaranteed within the distribution step of the hierarchical multi-level grid control strategy (see III. in Fig. 1). Nevertheless, the plausibility of the cost-structure can be investigated in more detail by comparing the costs for specific, representative operating points with the results of a cost-optimal provision of the corresponding IPFs. Based on the results an improved sampling of the FOR regarding monetarization aspects can be developed to reduce the gap.

## VII. Summary and Future Research Aspects

This paper continues research regarding the aggregation of ancillary service flexibility potentials within lower system levels at the vertical system interconnections in the context of hierarchical multi-(voltage-)level grid control strategies and especially TSO/DSO as well as DSO/DSO-cooperation. The main contribution is the monetarization of the active and reactive interconnection power flows (IPF) within the feasible operation region (FOR) by metadata from the particle swarm optimization (PSO). Advantages of the presented approach are the simple integration within the previous aggregation process without an extended sampling process and the independence from a specific local flexibility market concept. Furthermore, the possibility to consider a variety of different flexibility providing units (FPU) with individual economic interests for commodity and service prizes and even non-convex active and reactive power flexibility polygons is advantageous. Therefore, the aggregated flexibility potentials of multiple FPUs connected to a common bus are premonetarized by a cost-optimal flexibility disaggregation to the individual FPUs within a mixed-integer linear programming problem (MILP) prior to the FOR determination. Thereby, the cost-structure of the aggregated FPU can be determined in any level of detail. The metadata generated by the PSO during the sampling of the FOR and the resulting cost-structure of the FPUs are used to specify the service costs for a specific IPF and to monetarize the FOR.

The promising results of the case study are the basis for further extensions of the presented process and the application of other aggregation methods like random search and other metaheuristics. The potential extensions can be divided into three categories. The first category is coping with the implementation of further functionalities to the FOR aggregation method. Examples are the consideration of voltage dependencies of the FPU flexibility polygons and the integration of technology specific cost factors. The consideration of voltage dependencies at metaheuristics works analogously to the consideration of a three-dimensional FOR at the operational management of the higher-level system operator. The cost disaggregation within aggregated FPU-FORs can be simply modified to a mixed-integer quadratic programming (MIQP) problem for quadratic cost functions of the FPUs flexibility potentials. A better sampling of the FOR center is necessary in future assuming a small flexibility demand of the higher-level system operator. Another extension within the first category is the addition of further and the adaptation of current FPU flexibility polygons (e.g. bidirectional loading of electric vehicles) and the specification of more realistic investigation scenarios concerning the individual service prizes. For example, the shape of the flexibility polygons is influenced by partial load operation and on/off-states of the FPUs besides the regulating of the FPUs, which is already considered in this paper. Further developments require the extension of the aggregation methods to multiple vertical system interconnections.

The second category adapts FOR related research aspects from literature to implement them within the process presented in section III. Examples are the consideration of uncertainties due to forecast deviations, time constants for the FPU flexibility provision or the use of voltage controlling transformers as additional flexibility potential.

The third category deals with the further development and implementation of the monetarized FOR within the cascading process of a hierarchical multi-level grid control strategies. Therefore, the monetarized FOR at the DSO/DSO-interface has to be integrated within the operational management of the higher-level DSO. Beside a specification of an individual flexibility demand by the higher-level DSO the next step is the aggregation of the flexibilities for the TSO at the TSO/DSO-interface. For this, a determination of a cost function or cost factors within the FOR cost-structure is necessary. Based on this and the methods and investigations presented in this paper, the non-convex PQ-polygon of the FOR can be considered analogously to a FPU within the aggregation process at the TSO/DSO-interface.